\documentclass[twocolumn,american,superscriptaddress]{revtex4-1}
\usepackage[T1]{fontenc}
\usepackage[latin9]{inputenc}
\usepackage{amsmath}
\usepackage{graphicx}
\usepackage{wasysym}

\makeatletter

\newcommand*\LyXThinSpace{\,\hspace{0pt}}

\makeatother

\usepackage{babel}
\begin{document}

\title{Spin Hartree-Fock approach to quantum Heisenberg antiferromagnets
in low dimensions}

\author{A. Werth}

\affiliation{Instit\"ut fur Theoretische Physik, Universit\"at Frankfurt, Max-von-Laue
Strasse 1, 60438 Frankfurt, Germany}

\author{P. Kopietz}

\affiliation{Instit\"ut fur Theoretische Physik, Universit\"at Frankfurt, Max-von-Laue
Strasse 1, 60438 Frankfurt, Germany}

\affiliation{Department of Physics and Astronomy, University of California, Irvine,
CA, 92697, USA}

\author{O. Tsyplyatyev}

\affiliation{Instit\"ut fur Theoretische Physik, Universit\"at Frankfurt, Max-von-Laue
Strasse 1, 60438 Frankfurt, Germany}
\begin{abstract}
We construct a new mean-field theory for quantum (spin-1/2) Heisenberg
antiferromagnet in one (1D) and two (2D) dimensions using a Hartree-Fock
decoupling of the four-point correlation functions. We show that the
solution to the self-consistency equations based on two-point correlation
functions does not produce any unphysical finite-temperature phase
transition in accord with Mermin-Wagner theorem, unlike the common
approach based on the mean-field equation for the order parameter.
The next-neighbor spin-spin correlation functions, calculated within
this approach, reproduce closely the strong renormalization by quantum
fluctuations obtained via Bethe ansatz in 1D and a small renormalization
of the classical antiferromagnetic state in 2D. The heat capacity
approximates with reasonable accuracy the full Bethe ansatz result
at all temperatures in 1D. In 2D, we obtain a reduction of the peak
height in the heat capacity at a finite temperature that is accessible
by high-order $1/T$ expansions.
\end{abstract}

\date{\today}

\maketitle
Interest in the low-dimensional quantum antiferromagnets has been
revived in the last decades by the discovery of the high-$T_{c}$
superconductors, where the physics of the quantum spin fluctuations
on a square lattice was suggested to be the main mechanism behind
superconductivity~\cite{Anderson87}. More recently, the magnetic
properties of insulators such as $\textrm{Cs}_{2}\textrm{CuCl}_{4}$
\cite{Coldea03,Coldea09}, $\textrm{CsNiCl}_{3}$ \cite{Trudeau92},
$\textrm{Cs}_{2}\textrm{CuBr}_{4}$ \cite{Ono03}, where at low temperatures
a moderate degree of anisotropy (about $0.3-0.4$) makes their dimensionality
intermediate between one and two, have caused a new resurgence of
activity in this problem. In both cases the dimensionality is smaller
than three, where the classical long-range order in the ground state
was rigorously proven \cite{Dyson78,Kennedy88}, but is bigger than
strictly one, making the exact Bethe ansatz solution \cite{Gaudin_book}
inapplicable. Thus, a more accurate description of the effect of quantum
fluctuations is required, which become strong in reduced dimensions
and for the quantum spin $S=1/2$. 

A popular method to deal with low-dimensional quantum spin systems
is Takahashi's modified spin-wave theory that was quite successful
especially for ferromagnets, where it, for instance at low temperature,
reproduces correctly subleading terms of the free energy \cite{Takahashi87}
obtained using the thermodynamic Bethe ansatz approach \cite{Takahashi71,Gaudin71}.
Generally, this and other predictions of the Takahashi's theory are
almost equivalent to the Schwinger-boson mean-field theory formulated
by Arovas and Auerbach \cite{AurovasAuerbach88} and to the one-loop
renormalization group calculations \cite{Kopietz89}. However, at
high temperature the spin-wave result for the free energy is divergent
\cite{Takahashi87} disagreeing entirely with the high temperature
expansion in its limit of validity. In the antiferromagnetic case
predictions of the modified spin-wave theory are not as good for $S=1/2$.
In 1D, they lead to a gapful ground state and an exponential two-point
correlation function at zero temperature that deviates strongly from
the known from Bethe ansatz gapless ground state \cite{Gaudin_book}
and algebraic correlations \cite{Luther75,Kitanine99,Kitanine00,JES05}
at zero temperature. Also, in both 1D and 2D, there is a spurious
finite temperature phase transition within the spin-wave approach,
which is explicitly forbidden in these dimensions by the Mermin-Wagner
theorem \cite{MerminWagner66}. The latter problem stems from the
need of introducing two sublattices in the construction of the spin-wave
theory in the antiferromagnetic case \cite{AurovasAuerbach88,Takahashi89},
which is based on the simplest mean-field approximation using the
sublattice magnetization (a one-point correlation function) as the
order parameter and causes an order-disorder phase transition in all
dimensions that is not washed out by spin waves.

In this paper we construct an alternative mean-field approach for
the spin-1/2 antiferromagnet in 1D and 2D based on the decoupling
of the four-point correlation functions. The corresponding self-consistency
equations are derived using the Hartree-Fock decoupling for the Heisenberg
interactions and assuming the exclusive statistics of free magnons.
It recovers almost all effects of the strong renormalisation of the
classical spin picture in 1D at low temperature established by Bethe
ansatz, including the heat capacity and the static correlation functions,
with the most notable exception of the logarithmic contribution to
the magnetic susceptibility that is driven by the low-energy physics
of Luttinger liquid and requires taking into account even higher order
correlation functions. At high temperature our method recreates the
$1/T$ expansion and produces no phase transition at intermediate
temperatures. In 2D the same approach recovers only a small renormalisation
of the classical antiferromagnetic state \cite{Anderson52} in the
next-neighbor spin-spin correlation function and the high temperature
expansion, producing again no finite temperature phase transition.
The height of a smooth peak (instead of a transition) at an intermediate
temperature, for instance in the heat capacity, is reduced in 2D with
respect its value in 1D that is still accessible via high order $1/T$
expansion in 2D \cite{rushbrooke_book,Singh93} and is already captured
qualitatively on the level of the two-point correlation functions.
The biggest quantitative discrepancy of ignoring three- and higher-point
correlation functions occurs at intermediate temperatures and is of
the order of $20\%$ in 1D, where the thermodynamic quantities can
be calculated at arbitrary temperatures \cite{Takahashi73,Klumper93,Eggert94,Klumper00}
using the thermodynamic Bethe ansatz \cite{Takahashi71,Gaudin71}.

We study Heisenberg model for spin-1/2 in the presence of an external
magnetic field, $B$, in one ($D=1)$ and two ($D=2)$ dimensions,

\begin{equation}
H=B\sum_{\mathbf{r}}S_{\mathbf{r}}^{z}+\frac{J}{2}\sum_{\mathbf{r},\delta}\mathbf{S}_{\mathbf{r}}\cdot\mathbf{S}_{\mathbf{r}+\delta},\label{eq:Hr}
\end{equation}
where $J$ is the exchange energy, $S_{\mathbf{r}}^{z},S_{\mathbf{r}}^{\pm}=S_{\mathbf{r}}^{x}\pm iS_{\mathbf{r}}^{y}$
are the spin-1/2 operators at site $\mathbf{r}$, the sum over $\mathbf{r}$
runs over equidistant (square) lattice consisting of $N=L^{D}$ spins
in 1D (2D), and sum over $\delta$ runs over 2 or 4 nearest-neighbors
in the corresponding dimension. Below we impose periodic boundary
conditions, $\mathbf{S}_{\mathbf{r}+\mathbf{x}\left(\mathbf{y}\right)L}=\mathbf{S}_{\mathbf{r}}$,
restrict ourselves to the antiferromagnetic exchange energy, $J>0$,
and use the units where $g\mu_{B}=1$. 

Before proceeding with solving the model in Eq. (\ref{eq:Hr}), we
reduce the number of the spin components in it by utilizing the following
spin-1/2 identity $S_{\mathbf{r}}^{z}=S_{\mathbf{r}}^{+}S_{\mathbf{r}}^{-}-1/2$.
This turns the Zeeman term in the Hamiltonian into a quadratic form
and the $z$ component of the scalar product into a quartic form,
expressing Eq. (\ref{eq:Hr}) in terms of only $S_{\mathbf{r}}^{\pm}$
operators.

In the Fourier domain, $S_{\mathbf{r}}^{\pm}=N^{-1/2}\sum_{\mathbf{k}}S_{\mathbf{k}}^{\pm}e^{\pm i\mathbf{k}\cdot\mathbf{r}}$,
the resulting Hamiltonian becomes a sum of a quadratic and a quartic
form in the single spin operators,
\begin{multline}
H=\sum_{\mathbf{k}}\left(B-DJ+\varepsilon_{\mathbf{k}}\right)S_{\mathbf{k}}^{+}S_{\mathbf{k}}^{-}\\
+\frac{1}{N}\sum_{\mathbf{k}_{1}\mathbf{k}_{2}\mathbf{k}_{3}\mathbf{k}_{4}}\delta_{\mathbf{k}_{1}+\mathbf{k}_{3},\mathbf{k}_{2}+\mathbf{k}_{4}}\varepsilon_{\mathbf{k}_{3}-\mathbf{k}_{4}}S_{\mathbf{k}_{1}}^{+}S_{\mathbf{k}_{2}}^{-}S_{\mathbf{k}_{3}}^{+}S_{\mathbf{k}_{4}}^{-},\label{eq:Hk}
\end{multline}
where the dispersion is $\varepsilon_{\mathbf{k}}=J\sum_{\alpha}\cos k_{\alpha}$
, the sum $\sum_{\alpha}$ contains only one term $\alpha=x$ in 1D
and it runs over two spatial dimensions, $\sum_{\alpha=x,y}$, in
2D, and the sum over momentum, $\sum_{\mathbf{k}}$, also runs over
one ($k$) or two ($k_{x},k_{y}$) components of the wave vector in
the corresponding dimension.

In order to analyze the model in Eq. (\ref{eq:Hk}) we assume that
its eigenstates factorize in the momentum domain, \textit{i.e.} they
can approximated by product states of single magnon excitations in
the thermodynamic limit \cite{BKT}. At a finite temperature this
approach corresponds to writing down the following product density
matrix: $\rho=\prod_{\mathbf{k}}\left[m_{\mathbf{k}}\left|\mathbf{k}\left\rangle \right\langle \mathbf{k}\right|+\left(1-m_{\mathbf{k}}\right)\right]$,
where $\left|\mathbf{k}\right\rangle $ is a single magnon state at
a given $\mathbf{k}$, exclusive statistics for the states with different
$\mathbf{k}$ is implied \cite{nonexclusive}, $m_{\mathbf{k}}$ are
scalar parameters, and the normalisation is chosen as $\textrm{Tr}\rho=1$.
We believe that this density matrix gives a close enough approximation
to the many-magnon states. The expectation value of the Hamiltonian
in Eq. (\ref{eq:Hk}) with respect to this $\rho$ gives the energy
of the system, $E=\left\langle H\right\rangle $, as a function of
parameters $m_{\mathbf{k}}$, 
\begin{equation}
E=\sum_{\mathbf{k}}\left(B-DJ+\varepsilon_{\mathbf{k}}\right)m_{\mathbf{k}}-\frac{1}{N}\sum_{\mathbf{k}_{1}\mathbf{k}_{2}}\varepsilon_{\mathbf{k}_{1}-\mathbf{k}_{2}}m_{\mathbf{k}_{1}}m_{\mathbf{k}_{2}},\label{eq:Ek}
\end{equation}
where the contribution of the terms with $\mathbf{k}_{1}=\mathbf{k}_{2}$
in the second line vanishes in the $N\rightarrow\infty$ limit and
the average of an operator is $\left\langle \dots\right\rangle =\textrm{Tr}\left(\rho\dots\right)$.
The second term in Eq. (\ref{eq:Ek}) is equivalent to the Hartree-Fock
approximation to the quartic interaction term in Eq. (\ref{eq:Hk}),
$\left\langle S_{\mathbf{k}_{1}}^{+}S_{\mathbf{k}_{2}}^{-}S_{\mathbf{k}_{3}}^{+}S_{\mathbf{k}_{4}}^{-}\right\rangle \approx m_{\mathbf{k}_{1}}m_{\mathbf{k}_{3}}\delta_{\mathbf{k}_{1},\mathbf{k}_{2}}\delta_{\mathbf{k}_{3},\mathbf{k}_{4}}+m_{\mathbf{k}_{1}}\left(1-m_{\mathbf{k}_{2}}\right)\delta_{\mathbf{k}_{1},\mathbf{k}_{4}}\delta_{\mathbf{k}_{2},\mathbf{k}_{3}}$,
where the first term is the direct and the second is the exchange
part. The average of the operator $S_{\mathbf{k}}^{+}S_{\mathbf{k}}^{-}$
in the first term in Eq. (\ref{eq:Hk}) gives the scalar parameter
$\left\langle S_{\mathbf{k}}^{+}S_{\mathbf{k}}^{-}\right\rangle =m_{\mathbf{k}}$
that can be interpreted as a two-point correlation function. The inverse
Fourier transform gives the correlation function $\sum_{\mathbf{k}}e^{-i\mathbf{k}\cdot\mathbf{r}}m_{\mathbf{k}}/N=\left\langle S_{\mathbf{r}}^{+}S_{\mathbf{0}}^{-}\right\rangle $,
where $\mathbf{0}$ is a reference point on the lattice in 1D and
2D and the translational invariance of the model in Eq. (\ref{eq:Hr})
was used. 
\begin{figure}
\centering\includegraphics[width=1\columnwidth]{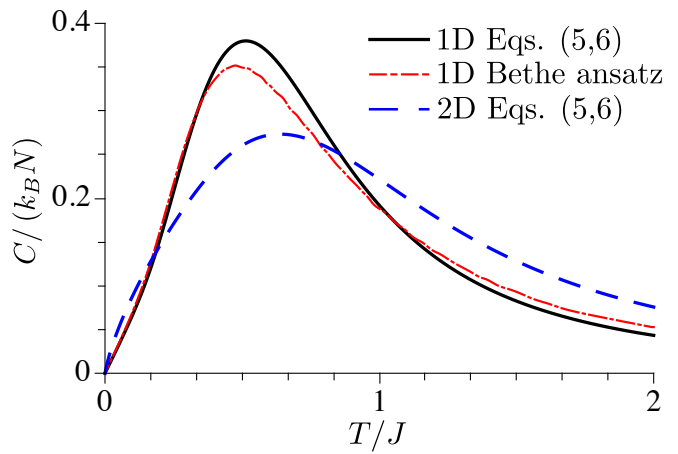}

\caption{\label{fig:c}Specific heat as a function of temperature at $B=0$
in 1D and 2D. The solid black and the blue dashed lines are obtained
solving the self-consistency Eqs. (\ref{eq:self_cons_s}, \ref{eq:self_cons_u})
and the heat capacity by means of Eq. (\ref{eq:c}). The red dash-dot
line is the exact result of thermodynamic Bethe ansatz calculation
in 1D from \cite{Klumper00}.}
\end{figure}

The values of the mean-field parameters $m_{\mathbf{k}}$ at a finite
temperature $T$ can be found in the usual way by minimizing the free
energy, $F=E-TS$, with respect to them. The energy $E$ is given
by Eq. (\ref{eq:Ek}) and the von Neumann entropy, $S=-k_{B}\textrm{Tr}\left(\rho\ln\rho\right)$,
is given by $-k_{B}\sum_{\mathbf{k}}\left[m_{\mathbf{k}}\ln m_{\mathbf{k}}+\left(1-m_{\mathbf{k}}\right)\ln\left(1-m_{\mathbf{k}}\right)\right]$,
where $k_{B}$ is the Boltzmann constant. Solving $\partial F/\partial m_{\mathbf{k}}=0$
we obtain the mean-field self-consistency equations as 
\begin{equation}
m_{\mathbf{k}}=\frac{1}{e^{\beta\left(B-DJ+\varepsilon_{\mathbf{k}}-\frac{2}{N}\sum_{\mathbf{k}'}\varepsilon_{\mathbf{k}-\mathbf{k}'}m_{\mathbf{k}'}\right)}+1},\label{eq:ak}
\end{equation}
where $\beta=1/\left(k_{B}T\right)$ is the inverse temperature. The
above is a large set of $N$ non-linear equations for the mean-field
parameters $m_{\mathbf{k}}$. However, the $m_{\mathbf{k}}$ enter
in the exponential function only under a sum. Thus, the number of
the independent non-linear equations can be reduced greatly. We introduce
$1+D$ extensive variables as $s=\sum_{\mathbf{k}}m_{\mathbf{k}}/N-1/2$
and $u_{\alpha}=-\sum_{\mathbf{k}}m_{\mathbf{k}}\cos k_{\alpha}/N+1/2$,
where $u_{\alpha}=u$ is a scalar in 1D and $u_{\alpha}=\left(u_{x},u_{y}\right)$
is a vector in 2D. Substituting these definitions into Eq.~(\ref{eq:ak})
we express its the right-hand-side in terms of only $s$ and $u_{\alpha}$
and then substituting the resulting expressions for $m_{\mathbf{k}}$
back into the definitions for $s$ and $u_{\alpha}$ we rewrite Eq.(\ref{eq:ak})
as a set of only $1+D$ independent equations,

\begin{align}
s & =\int\frac{d^{D}k}{\left(2\pi\right)^{D}}\frac{1}{e^{\beta\left(B+2DJs+2J\sum_{\alpha}u_{\alpha}\cos k_{\alpha}\right)}+1}-\frac{1}{2},\label{eq:self_cons_s}\\
u_{\alpha} & =\frac{1}{2}-\int\frac{d^{D}k}{\left(2\pi\right)^{D}}\frac{\cos k_{\alpha}}{e^{\beta\left(B+2DJs+2J\sum_{\alpha}u_{\alpha}\cos k_{\alpha}\right)}+1},\label{eq:self_cons_u}
\end{align}
where the sum over $\mathbf{k}$ was turned into an integral in the
thermodynamic limit as $\sum_{\mathbf{k}}/N\rightarrow\int d^{D}k/\left(2\pi\right)^{D}$
\cite{JW}. Here the parameter $s$ gives the average magnetization
per spin as $\sum_{\mathbf{r}}\left\langle S_{\mathbf{r}}^{z}\right\rangle /N=s$
and the parameter $u_{\alpha}$ is related to the kinetic energy of
magnons.

There is only one non-trivial solution of Eqs. (\ref{eq:self_cons_s},
\ref{eq:self_cons_u}). Let us analyze it at $B=0$. At zero temperature
the integrands are proportional to the Heaviside step function, $\lim_{\beta\rightarrow\infty}\left[\exp\left(\beta x\right)+1\right]^{-1}=\Theta\left(-x\right)$,
then the integrals can be calculated explicitly, and we obtain $s=0$
(unpolarized ground state) and $u_{x}=u_{y}=1/2+D/\pi^{D}$. On the
other hand, at high temperature, the exponential expands into a Taylor
series in $\beta\ll1$ up to the leading order as $\left[\exp\left(\beta x\right)+1\right]^{-1}=1/2+O\left(\beta\right)$
and we get $s=0$ and $u_{x}=u_{y}=1/2$. At intermediate temperatures
the equations can be solved numerically.

The thermodynamic quantities can be expressed through solutions of
Eqs. (\ref{eq:self_cons_s}, \ref{eq:self_cons_u}) at different temperatures
and magnetic fields. The energy in Eq. (\ref{eq:Ek}) can be written
as a function of $s$ and $u_{\alpha}$ using their definitions in
terms of $m_{\mathbf{k}}$: $E=N\left(Bs+DJs^{2}-J\sum u_{\alpha}^{2}\right)$.
From which, using the basic definition of the heat capacity we obtain

\begin{equation}
\frac{C}{N}=\frac{1}{N}\frac{\partial E}{\partial T}=\left(B+2DJs\right)\frac{\partial s}{\partial T}-2J\sum_{\alpha}u_{\alpha}\frac{\partial u_{\alpha}}{\partial T}.\label{eq:c}
\end{equation}
The temperature dependence of $C$ in 1D and in 2D (at $B=0$) is
plotted in Fig.~\ref{fig:c}. In 1D we can compare our result with
the full quantum mechanical result obtained via the thermodynamic
Bethe ansatz machinery \cite{Takahashi71,Gaudin71} in \cite{Takahashi73,Klumper93,Klumper00}.
Up to the intermediate temperatures Eqs. (\ref{eq:self_cons_s}, \ref{eq:self_cons_u})
agree quite well with it including the linear dependence of $C$ at
low temperatures. Eqs. (\ref{eq:self_cons_s}, \ref{eq:self_cons_u})
also reproduce the correctly coefficient of the leading term of the
$1/T$ expansion at high temperatures. However, in the intermediate
temperature region, from $T\apprge J/2$, the difference, see the
black solid and the red dash-dotted lines in Fig.~\ref{fig:c}, is
still appreciable, up to $20\%$. In 2D the available high order $1/T$
expansion \cite{rushbrooke_book} covers a significant temperature
range down to the peak, which amplitude is reduced with respect to
the 1D case. The result of solving Eq. (\ref{eq:self_cons_s},\ref{eq:self_cons_u}),
the blue dashed line in Fig.~\ref{fig:c}, gives about the same discrepancy
of up to $20\%$ with \cite{rushbrooke_book} in the intermediate
temperature region.
\begin{figure}
\centering\includegraphics[width=1\columnwidth]{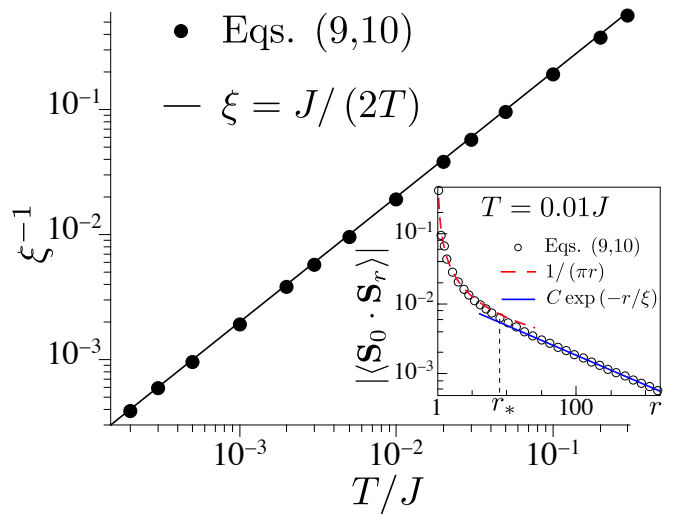}

\caption{\label{fig:1D_corr}The static correlation function $\left\langle \mathbf{S}_{0}\cdot\mathbf{S}_{r}\right\rangle $
in 1D evaluated by solving the self-consistency equation in Eqs. (\ref{eq:self_cons_s},
\ref{eq:self_cons_u}) and using Eqs. (\ref{eq:corr_func},\ref{eq:Ir}).
The main plot is the temperature dependence of the correlation length
$\xi$ in the exponential regime $r>r_{*}$ obtained numerically (full
circles) and the $1/T$ result of Bethe ansatz (black line). The inset
is the correlation function for an intermediate temperature $T=0.01J$
obtained numerically (open circles); the zero temperature result,
$1/r$, is valid in the short-range region $r<r_{*}$ (dashed red
line) and a exponential, $\exp\left(-r/\xi\right)$, is valid in the
long-range region $r>r_{*}$ (solid blue line). }
\end{figure}

This discrepancy can be understood in term of improving approximate
description of the Heisenberg model by taking into account higher
order correlation functions. The usual way of introducing the mean-field
approximation to the model in Eq. (\ref{eq:Hr}) is retaining only
the one-point correlation functions, $\left\langle S_{\mathbf{r}}^{z}\right\rangle =\pm m$,
where $m$ is the order (a single mean-field) parameter and $\pm$
describes even/odd sublattice of the antiferromagnet. Neglecting the
quadratic terms in fluctuations around $\left\langle S_{\mathbf{r}}^{z}\right\rangle $
(and assuming that $\left\langle S_{\mathbf{r}}^{\pm}\right\rangle =0$)
in Eq. (\ref{eq:Hr}), the usual self-consistency equations are the
same for each sublattice for $B=0$,

\begin{equation}
2m=\tanh\left(\beta DJm\right).\label{eq:a}
\end{equation}
This equation does predict the antiferromagnetic order at $T=0$,
but it also introduces an erroneous phase transition at a finite $T$
in low dimensions, which is explicitly forbidden by the Mermin-Wagner
theorem \cite{MerminWagner66}. In the present work we take into account
two-point correlation function solving the self-consistency equations
in Eq. (\ref{eq:ak}) for $N$ mean-field parameters $m_{\mathbf{k}}$.
This approach contains more information about the quantum fluctuations,
which play a stronger role in low dimensions, improving the approximation
qualitatively, \textit{i.e.} not introducing a finite $T$ phase transition,
and quantitatively, as illustrated in 1D by comparison with the Bethe
ansatz in Fig. \ref{fig:c}. An approach that accounts for higher
than two-point correlation functions would improve the accuracy even
further.

Another thermodynamic quantity that is of interest in magnets is the
magnetic susceptibility, $\chi=\partial\left(\sum_{\mathbf{r}}\left\langle S_{\mathbf{r}}^{z}\right\rangle \right)/\partial B$.
Using as before the identity $S_{\mathbf{r}}^{z}=S_{\mathbf{r}}^{+}S_{\mathbf{r}}^{-}-1/2$
and the definition of $s$ in terms of $m_{\mathbf{k}}$, we obtain
$\chi=N\partial s/\partial B.$ The temperature dependence of this
result at $B=0$ shows a better (in comparison with the more crude
approximation in Eq. (\ref{eq:a})) agreement with the full Bethe
ansatz calculation \cite{Eggert94,Klumper00}, both quantitatively
and qualitatively. However, there are larger deviations at small temperatures,
unlike for the heat capacity, due to the logarithmic corrections \cite{Eggert94}.
They are essentially an effect of Luttinger physics manifesting hydrodynamic
modes which are not captured on the level of the two-point correlation
functions in Eq. (\ref{eq:ak}).

The static correlation functions can also be calculated in terms of
the solutions of Eqs. (\ref{eq:self_cons_s},\ref{eq:self_cons_u}).
Expressing the operator $\mathbf{S}_{\mathbf{0}}\cdot\mathbf{S}_{\mathbf{r}}$
in the Fourier domain and evaluating its finite temperature average
using $\rho$ in the same way as in the calculation of the energy
of the system in Eq. (\ref{eq:Ek}) we obtain

\begin{equation}
\left\langle \mathbf{S}_{\mathbf{0}}\cdot\mathbf{S}_{\mathbf{r}}\right\rangle =s^{2}+I\left(\mathbf{r}\right)\left[1-I\left(\mathbf{r}\right)\right]\label{eq:corr_func}
\end{equation}
where
\begin{equation}
I\left(\mathbf{r}\right)=\int\frac{d^{D}k}{\left(2\pi\right)^{D}}\frac{\cos\left(\mathbf{k}\cdot\mathbf{r}\right)}{e^{\beta\left(B+2DJs+2J\sum_{\alpha}u_{\alpha}\cos k_{\alpha}\right)}+1}.\label{eq:Ir}
\end{equation}
Here $m_{\mathbf{k}}$ were expressed through $s$ and $u_{\alpha}$
using their definitions above. For the next-neighbor correlation function
the integral in Eq. (\ref{eq:Ir}) simplifies even further using Eq.
(\ref{eq:self_cons_u}): $I\left(1\right)=1/2-u$ and $I\left(\mathbf{x}\right)=1/2-u_{x}$
in the corresponding dimension. At zero temperature we can substitute
the already obtained solutions of Eqs.~(\ref{eq:self_cons_s},\ref{eq:self_cons_u}),
$s=0$ and $u_{x}=u_{y}=1/2+D/\pi^{D}$, directly. In 1D, where quantum
fluctuations play a significant role, we obtain $\left\langle \mathbf{S}_{0}\cdot\mathbf{S}_{1}\right\rangle =-0.4196\dots$
that is close to the full Bethe ansatz result $\left\langle \mathbf{S}_{0}\cdot\mathbf{S}_{1}\right\rangle =-0.4431\dots$
\cite{Orbach58}. In two dimension we obtain $\left\langle \mathbf{S}_{\mathbf{0}}\cdot\mathbf{S}_{\mathbf{x}}\right\rangle =-0.2437\dots$
that is close to the value of $-1/4$ for the classical antiferromagnet,
with only a small reduction due to quantum fluctuations \cite{Anderson52}.

Beyond the next-neighbor the integral in Eq. (\ref{eq:Ir}) needs
to be calculated explicitly. At $T=0$ it gives $I\left(r\right)=\sin\left(\pi r/2\right)/\left(\pi r\right)$
in 1D resulting in the correlation function $\left\langle \mathbf{S}_{0}\cdot\mathbf{S}_{r}\right\rangle =\sin\left(\pi r/2\right)/\left(\pi r\right)$
at $r\gg1$. This $1/r$ behavior coincides with the prediction of
a Gaussian conformal field theory \cite{Luther75} that was confirmed
by direct Bethe ansatz calculation of the corresponding form-factors
\cite{Kitanine99,Kitanine00,JES05}. At a finite $T>0$ numerical
solution of Eqs.~(\ref{eq:self_cons_s},\ref{eq:self_cons_u}) and
numerical evaluation of the integral in Eq. (\ref{eq:Ir}) give an
exponential behavior $\left|\left\langle \mathbf{S}_{0}\cdot\mathbf{S}_{r}\right\rangle \right|\propto\exp\left(-r/\xi\right)$
at large distances, see the fit in the inset in Fig. \ref{fig:1D_corr},
where the correlation length in 1D also obtained by fitting is an
algebraic function of temperature,
\begin{equation}
\xi=\frac{J}{2T},
\end{equation}
see the main part in Fig. \ref{fig:1D_corr}. This coincides with
the $1/T$ behavior obtained using the thermodynamic Bethe ansatz
approach \cite{Klumper93}. The exponential behavior crosses over
into the power law at a finite range $r_{*}$, see the inset in Fig.
\ref{fig:1D_corr}, which value changes smoothly from $r_{*}=\infty$
at $T=0$ to $r_{*}\approx0$ at $T\simeq J$. In 2D the integral
in Eq. (\ref{eq:Ir}) gives $I\left(r\mathbf{x}\right)=-2\sin\left(\pi r/2\right)/\left(\pi r\right)^{2}$
at zero temperature and the $\left\langle \mathbf{S}_{\mathbf{0}}\cdot\mathbf{S}_{\mathbf{x}r}\right\rangle =-2\sin\left(\pi r/2\right)/\left(\pi r\right)^{2}$
correlation function at $r\gg1$. At finite temperature the correlation
length in the two-dimensional antiferromagnet is known to have an
exponential dependence on temperature, $\xi\propto\exp\left(\textrm{const}/T\right),$
\cite{Chakravarty88}. Numerically we find that the result of Eqs.
(\ref{eq:self_cons_s},\ref{eq:self_cons_u},\ref{eq:Ir}) is consistent
with \cite{Chakravarty88} at a small temperature range below $T\simeq J$,
which is still accessible due to not so large values of $r_{*}$ at
relatively not so low temperatures.

In conclusion, we have constructed a new mean-field approach based
on two-point correlation functions for spin-1/2 antiferromagnet in
1D and 2D, for which the effect of quantum fluctuations is the strongest.
Solutions of the corresponding self-consistency equations recover
the strong renormalisation of the classical spin picture in 1D, established
by Bethe ansatz, and only a small corrections to the classical antiferromagnet
in 2D. This approach produces no finite temperature phase transitions
in accord with the Mermin-Wagner theorem and the $1/T$-expansion
at high temperature in $D=1$ and $D=2$. The biggest quantitative
discrepancy of ignoring three- and higher-point correlation functions
occurs at intermediate temperatures and is up to $\sim20\%$ that
can be assessed in 1D, where the thermodynamic quantities can be calculated
at arbitrary temperatures using the thermodynamic Bethe ansatz. The
controversy about the effect of dimensionality in the anisotropic
2D quantum antiferromagnets, \textit{e.g.} $\textrm{Cs}_{2}\textrm{CuCl}_{4}$
(the ratio of the exchange constants is $J_{\perp}/J_{\parallel}\simeq0.33$)
for which neutron scattering shows both signatures of one-dimensional
physics \cite{Coldea97} and a dispersion in the perpendicular direction
\cite{Coldea01}, can be explained here as a dimensional crossover,
where strong effects of quantum fluctuations in 1D disappear smoothly
as the coupling between the chains is increased.

We acknowledge financial support by the DFG through SFB/TRR 49. PK
acknowledges the hospitality of the Department of Physics and Astronomy
of the University of California, Irvine, where a part of this work
was done.


\begin{thebibliography}{10}
\bibitem{BKT}This assumption restricts the possibility of a Berezinskii\textendash Kosterlitz\textendash Thouless
physics in 2D (analogously to the classical XY model \cite{Berezinskii71,KosterlitzThouless73}),
which for the quantum Heisenberg model however may be washed out by
the quantum fluctuations of spin-1/2.
\bibitem{nonexclusive}Assumption of non-exclusive statistics here
does not recover effects of the strong quantum fluctuations in 1D
and the high-$T$ limit.
\bibitem{JW}Similar equations were obtained using Jordan-Wigner transformation
strictly in 1D in \cite{Bulaevskii63}. A renormalized single-particle
like dispersion in the exponent of Eqs.~(\ref{eq:self_cons_s},\ref{eq:self_cons_u})
also reproduces qualitatively \cite{OT14} the Bethe ansatz result
and gives the leading contribution to observables in the thermodynamic
limit at high energy \cite{OT15}.
\end{thebibliography}
\end{document}